\documentclass{jps-cp}
\usepackage{txfonts} 

\title{Spread of infectious diseases: Effects of the treatment of population}

\author{Kazue \textsc{Kudo} and Kanako \textsc{Mizuno}}

\inst{Department of Computer Science, Ochanomizu University, 
2-1-1 Ohtsuka, Bunkyo-ku, Tokyo 112-8610, Japan}

\email{kudo@is.ocha.ac.jp}

\recdate{October 17, 2016}

\abst{In a metapopulation network, infectious diseases spread widely
because of the travel of individuals. In the present study, we consider a modified metapopulation
Susceptible-Infected-Removed (SIR) model with a latent period, which we
call the SHIR model. In the SHIR model, an infectious period is divided 
into two stages. In the first stage, which corresponds to the latent period,
infectious individuals can travel. However, in the second stage,
the same individuals cannot travel since they are seriously ill. 
Final size distributions of the
metapopulation SIR and SHIR models are simulated with two different
methods and compared. 
In Monte Carlo simulations, in which the population is treated as an
integer, the distributions show similar behavior.
However, in reaction-diffusion systems, in which the population is treated as
a real number, the final size distribution of
the SHIR model has a discontinuous jump, and that of the SIR model
shows a continuous transition. 
The discontinuous jump is found to be an artifact that occurs owing to an
inappropriate termination condition.} 

\kword{epidemic model, metapopulation network}

\usepackage{amsmath}	
\begin{document}
\maketitle

\section{Introduction}

The spread of infectious diseases is often modeled on a complex network. 
If the network represents a human network, each node corresponds to an
individual. Transmission occurs between connected nodes.
The epidemic condition depends on the structure of the network as well as
the transmission and recovery rates. 
The impact of an epidemic is characterized by the basic
reproduction number $R_0$, which is the expected number of infections
caused by a typical infectious individual in a completely susceptible
population~\cite{Anderson91,Ma}.
$R_0=1$ represents the epidemic threshold.
In the Susceptible-Infected-Removed (SIR) model,
$R_0$ is given simply by the ratio of the transmission rate to
the recovery rate
when the network is a complete graph, or under a mean-field
approximation.
However, when the network is a complex network, e.g., a scale-free
network, the epidemic threshold depends also on the average connectivity
between the nodes~\cite{May}. 

Human mobility is also an important factor behind the spread of infectious
diseases. Metapopulation models are often used to implement the movement of
individuals.
In a metapopulation network, each node corresponds to a subpopulation or 
patch, which contains individuals. 
In metapopulation SIR models, 
while infection dynamics occur only within each patch,
individuals can move between neighboring 
patches~\cite{Keeling,Cross,Colizza07,Colizza08}. 
The epidemic threshold in each patch is given by the basic reproduction
number $R_0$.  
The epidemic in a metapopulation system is characterized by not only the
epidemic 
threshold but also the global invasion threshold, which depends on 
the mobility rate and network structure, as well as the basic
reproduction number~\cite{Cross,Colizza07,Colizza08}.

In the current study, we consider a modified SIR model with a
latent period, which we call the SHIR model,
on a metapopulation network~\cite{Mizuno15}.
In the SHIR model, an infectious period is divided into two
stages: an infected stage (denoted by ``H''), which corresponds to the
latent period,  
and a seriously-ill stage (denoted by ``I''), in which the individual
is infected and cannot move to another patch.
In Ref.~\cite{Mizuno15}, unexpected results were reported; 
a discontinuous jump appeared in the final size distribution, which
expresses the number of recovered individuals in the SHIR model,
and the global invasion threshold was almost
independent of the  mobility rate.
In this paper, we explain the reason behind those results, based on
numerical simulations performed with two different methods.
 
The rest of the paper is organized as follows.
The metapopulation SIR and SHIR models are introduced
in Sec.~\ref{sec:2}. 
In Sec.~\ref{sec:3}, we demonstrate the final size
distributions simulated with two different methods.
The time evolution of the number of infected individuals is also shown to
explain the discontinuous jump in the final size distribution.
Finally, conclusions are presented in Sec.~\ref{sec:4}.

\section{\label{sec:2} Models}

First, we introduce a metapopulation SIR model in which individuals can
move between connected patches.  
In the metapopulation SIR model, each patch can contain many
individuals, and each individual is in one of the three 
states: Susceptible,  Infected, or Recovered. 
Infection occurs in each patch,
and a disease is spread by the movement of individuals.
Each individual can move from one patch to another that is connected
with it.
If we assume that the mobility rate $w$ is the same in the entire network,
the time evolution of $S_i$, $I_i$, and $R_i$ (the numbers of
susceptible, infected, and recovered individuals in the $i$-th patch, respectively) 
is described by the following equations.
\begin{subequations}
\begin{align}
 \frac{dS_i}{dt} &= - \beta\frac{S_i I_i}{N_i}
 + w\sum_{j\in\partial i}(S_j - S_i),  
\label{eq:SIR.S}\\
 \frac{dI_i}{dt} &= \beta\frac{S_i I_i}{N_i} - \gamma I_i 
+ w\sum_{j\in\partial i}(I_j - I_i), 
\label{eq:SIR.I}\\
 \frac{dR_i}{dt} &= \gamma I_i 
+ w\sum_{j\in\partial i}(R_j - R_i). 
\label{eq:SIR.R}
\end{align}
\label{eq:SIR}%
\end{subequations}
Here, $\beta$ and $\gamma$ are the transmission and recovery rates,
respectively, and $N_i=S_i+I_i+R_i$ is the total number of individuals
in the $i$-th patch.  
The notation $\partial i$ represents the set of neighbors of the $i$-th
patch.  
In other words, the summations are taken over all the patches that are 
connected to the $i$-th patch. 

Next, we modify the metapopulation SIR model to take into account a latent
period. 
If a person were sick, he/she probably would not go out. Then, infected
people who can travel are considered to be the ones who are in a latent
period.   
Under this assumption, an infectious period is divided into two stages:
infected (in a latent period) and seriously-ill stages.
In the first stage, an individual can infect other individuals
and travel between neighboring patches.
However, in the second stage, an individual cannot travel.
In the modified model, the time evolution of $S_i$, $H_i$ , $I_i$ and 
$R_i$, (i.e., the number of susceptible, infected, seriously-ill, and recovered
individuals in the $i$-th patch) are 
described by the following equations.
\begin{subequations}
\begin{align}
 \frac{dS_i}{dt} &= - \beta\frac{S_i(H_i+I_i)}{N_i}
 + w\sum_{j\in\partial i}(S_j - S_i),  
\label{eq:SHIR.S}\\
 \frac{dH_i}{dt} &= \beta\frac{S_i(H_i+I_i)}{N_i} - \mu H_i 
+ w\sum_{j\in\partial i}(H_j - H_i), 
\label{eq:SHIR.H}\\
 \frac{dI_i}{dt} &= \mu H_i - \gamma I_i, 
\label{eq:SHIR.I}\\
 \frac{dR_i}{dt} &= \gamma I_i 
+ w\sum_{j\in\partial i}(R_j - R_i). 
\label{eq:SHIR.R}
\end{align}
\label{eq:SHIR}%
\end{subequations}
Here, $\mu$ is the transition rate from the infected to the seriously-ill
stage, and $N_i=S_i+H_i+I_i+R_i$.
We call this model the SHIR model.
This model is different from the SEIR model~\cite{Schwartz}, which is an
epidemic model in which a latent period is expressed as
an ``Exposed'' state. The exposed individual is infected,
but does not infect others.
However, the SHIR model as well as the SEIR model is one of the generalized
SIR models that include multiple infectious stages~\cite{Ma}.

The basic reproduction number $R_0$ in the standard
SIR model (without mobility) is given by $\beta/\gamma$, and
that in the generalized SIR model that
includes $n$ infectious stages is given by
\begin{equation}
 R_0=\sum_{i=1}^n\frac{\beta_i}{\gamma_i},
\label{eq:multi_R0}
\end{equation}
where $\beta_i$ is the transmission rate in 
the $i$-th infectious stage, and $1/\gamma_i$ is the mean duration
 an individual spends in the $i$-th infectious stage~\cite{Ma,Hyman}.
In Eq.~(\ref{eq:SHIR}),
$\beta_1=\beta_2=\beta$, $\gamma_1=\mu$ and
$\gamma_2=\gamma$. Thus, we find 
$R_0=\beta(\mu+\gamma)/(\mu\gamma)$ for the SHIR model (without
mobility)~\cite{Mizuno15}. 

\section{\label{sec:3} Final Size Distributions}

In this section, we demonstrate the final size distributions of the
metapopulation SIR and SHIR models.
The final size of an epidemic is the cumulative number of infected
individuals, which is equal to the total number of recovered individuals 
at the end of the epidemic.
The attack ratio, which is the normalized final size, is given by the
ratio of the number of recovered individuals to the total number of
individuals. 

The metapopulation network we use is a scale-free network of $500$ nodes, whose degree
distribution is $P(k)\sim k^{-\Gamma}$ with $\Gamma=3.5$.
Initially, one individual is infected, and the others are susceptible.
The initial number of individuals at each node is taken as
$\bar{N}k/\langle k \rangle$, where $\langle k \rangle$ is the average
degree, with $\bar{N}=100$, unless otherwise specified. 
In the numerical simulations below, we fix parameters as $\gamma=0.25$
for the SIR and $\gamma=\mu=0.5$ for the SHIR models. 
Then, the epidemic threshold $R_0=1$ corresponds to $\beta=0.25$ in both
the models. 

\begin{figure}[tbh]
\includegraphics[width=8cm]{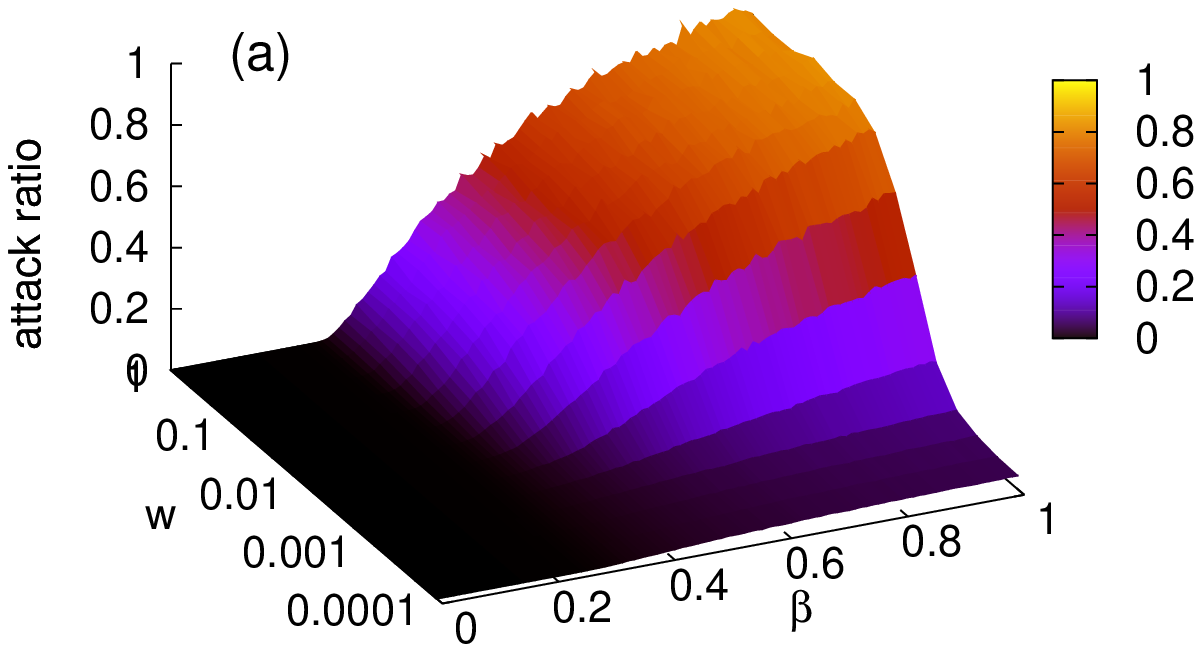}
\includegraphics[width=8cm]{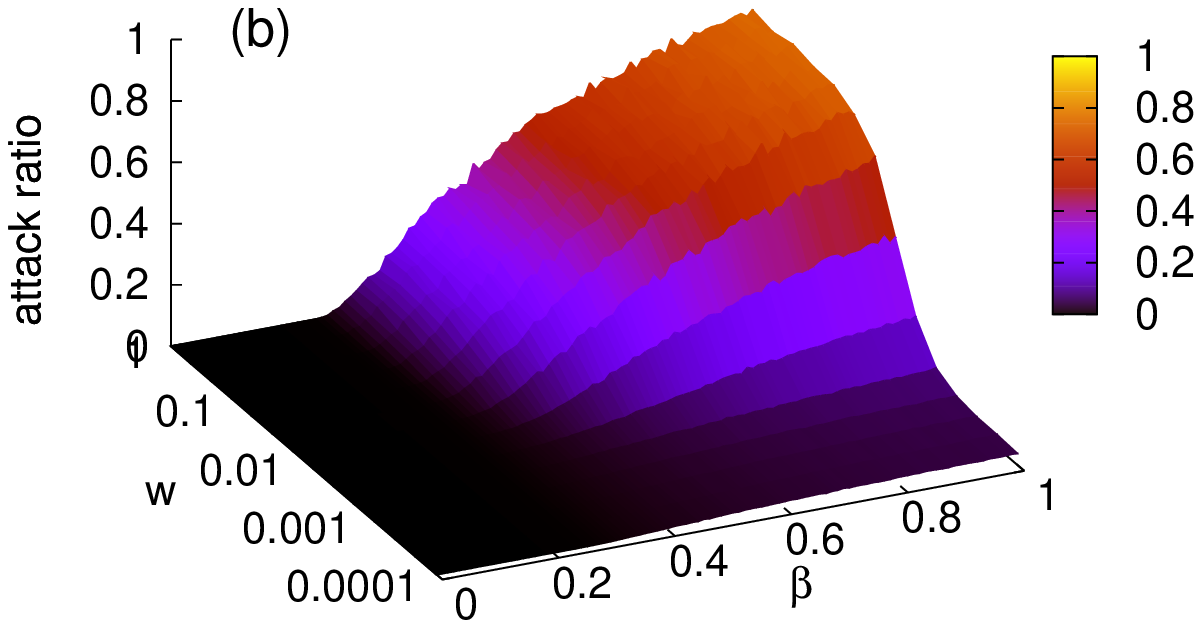}
\caption{Final size distributions simulated with a Monte Carlo method for (a)
 a metapopulation SIR model and (b) a metapopulation SHIR model. $w$ and
 $\beta$ are the mobility and transmission rates, respectively.}
\label{f1}
\end{figure}

We perform numerical simulations with two different methods. 
The first one is the Monte Carlo method. One Monte Carlo step consists
of transmission and traveling processes. 
In the case of the SIR model, in the $i$-th patch,
a susceptible individual becomes infected with probability 
$1-(1-\frac{\beta}{N_i}\tau)^{I_i}$, 
where $\tau$ is the time scale, 
and each infected individual recovers with probability
$\gamma\tau$~\cite{Colizza07}. 
In the SHIR model, a susceptible individual becomes infected with
probability  $1-(1-\frac{\beta}{N_i}\tau)^{H_i+I_i}$, 
an infected individual turns  seriously ill with probability $\mu\tau$, 
and a seriously-ill individual recovers with probability $\gamma\tau$. 
Each individual who can travel, moves from one patch with degree $k$ to
another neighboring patch with probability $w\tau/k$.
In Monte Carlo simulations, $S_i$, $H_i$, $I_i$, and $R_i$ are treated as
integers. 
In one Monte Carlo step, in which the time scale is taken as $\tau=1$, 
the transmission process is followed by the traveling process. 
States of the individuals are updated after each process.
Simulation continues as long as $\sum_i I_i \ge 1$ for the SIR model
and $\sum_i (H_i + I_i) \ge 1$ for the SHIR model.
The final size distributions for the metapopulation SIR and SHIR models are
shown in Figs.~\ref{f1}(a) and \ref{f1}(b), respectively. 
In both the distributions, the attack ratio starts to increase near
$\beta=0.25$, which corresponds to the epidemic threshold of $R_0=1$.
When the mobility rate is small, the attack ratio is also small in both 
Figs.~\ref{f1}(a) and \ref{f1}(b). 

\begin{figure}[tbh]
\includegraphics[width=8cm]{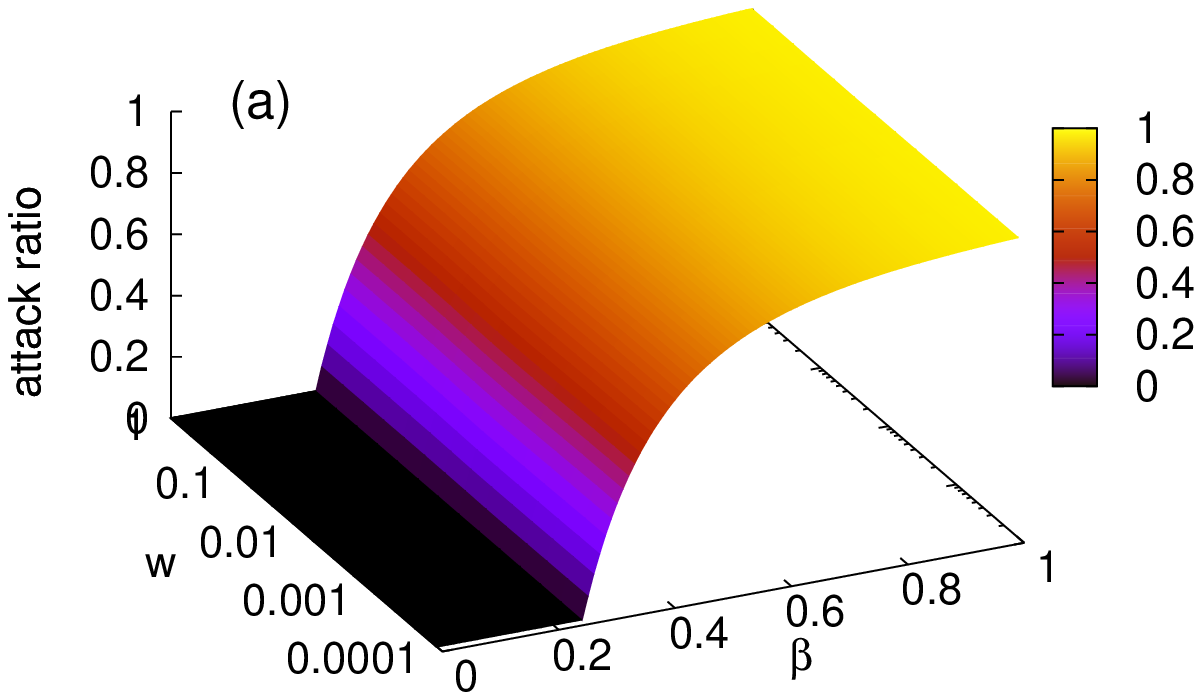}
\includegraphics[width=8cm]{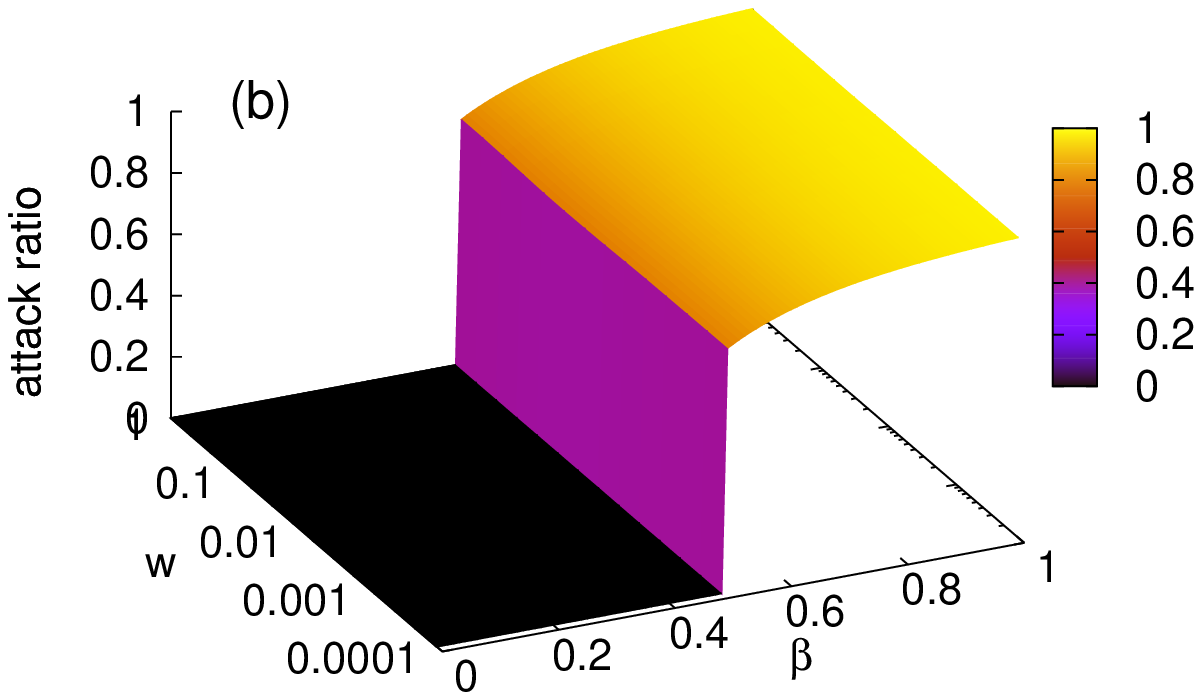}
\caption{Final size distributions in reaction-diffusion systems for (a)
 a metapopulation SIR model and (b) a metapopulation SHIR model. $w$ and
 $\beta$ are mobility and transmission rates, respectively.}
\label{f2}
\end{figure}

The second numerical method involves solving the reaction-diffusion equations
(\ref{eq:SIR}) and (\ref{eq:SHIR}) numerically. We employ the 4th
Runge-Kutta method for time evolution.
In this method, $S_i$, $H_i$, $I_i$, and $R_i$ are treated as real
numbers. 
In Ref.~\cite{Mizuno15}, termination conditions, which are conditions
to stop a simulation, were $\sum_i I_i <1$
for the SIR model and $\sum_i (H_i + I_i) <1$ for the SHIR model.
Figures~\ref{f2}(a) and \ref{f2}(b) show the final size distributions under these termination conditions for
the metapopulation SIR and SHIR models, respectively. 
In Fig.~\ref{f2}(a), we see a continuous transition at $\beta=0.25$
($R_0=1$). However, in Fig.~\ref{f2}(b), a discontinuous jump appears
at $\beta=0.5$. 
Another important point is that the final size distributions are
not related to the mobility rate $w$. 
These results are almost the same as the ones reported in Ref.~\cite{Mizuno15}. 

The reason why the final size distribution is independent of $w$ is owing to  the fact that the second method is a type of a 
mean-field approach in which the network structure is irrelevant.
In other words, the entire metapopulation network can be regarded as a
large single population in the mean-field approach.
In such a case, attack ratio does not depend on the mobility rate,
although the time scales for the epidemic to spread and cease depend on the
mobility rate.
By contrast, in Monte Carlo simulations, the final size distributions 
are strongly dependent on the mobility rate. Furthermore, in this case, the network structure is
relevant as well.
When the mobility rate is very small, the epidemic cannot spread to
other patches even if the transmission rate is large, although a local
outbreak may occur. Since a large fraction of patches stay uninfected for
a small mobility rate, the attack ratio is small even for a large
transmission rate. 

\begin{figure}[tbh]
\includegraphics[width=8cm]{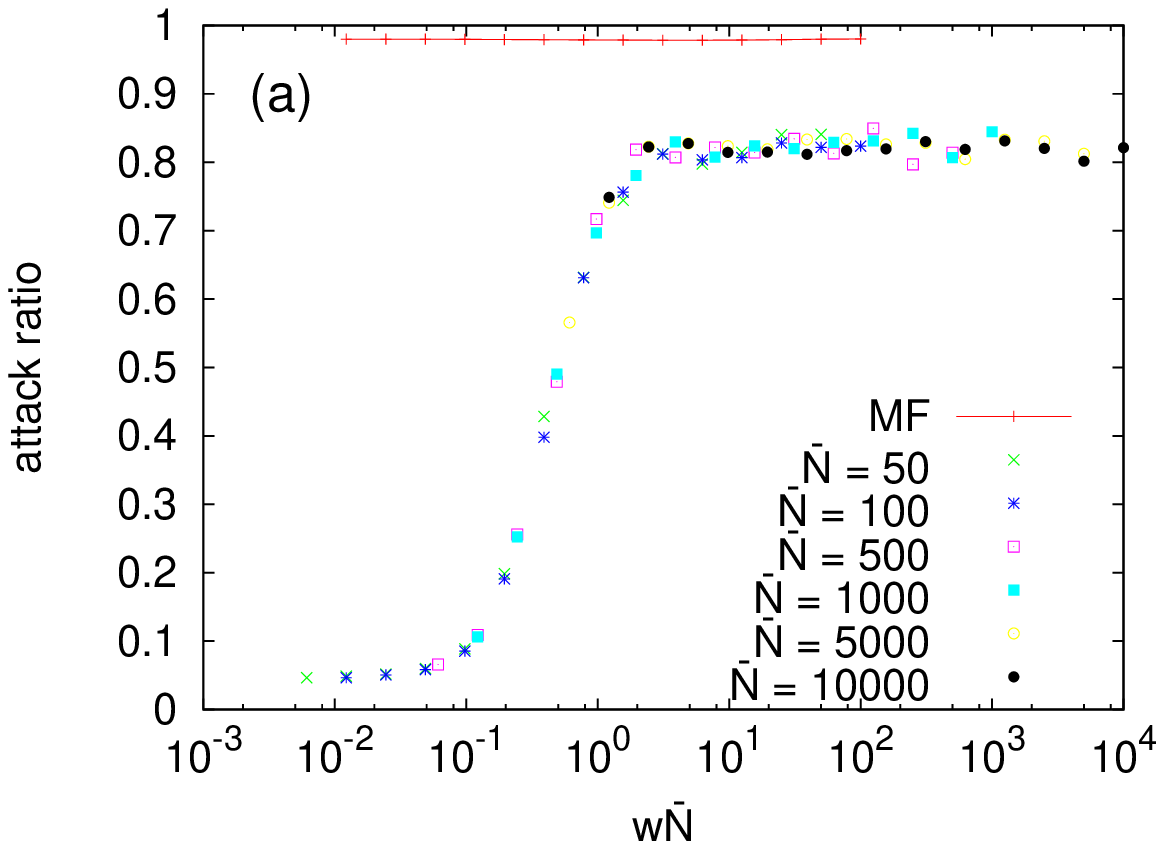}
\includegraphics[width=8cm]{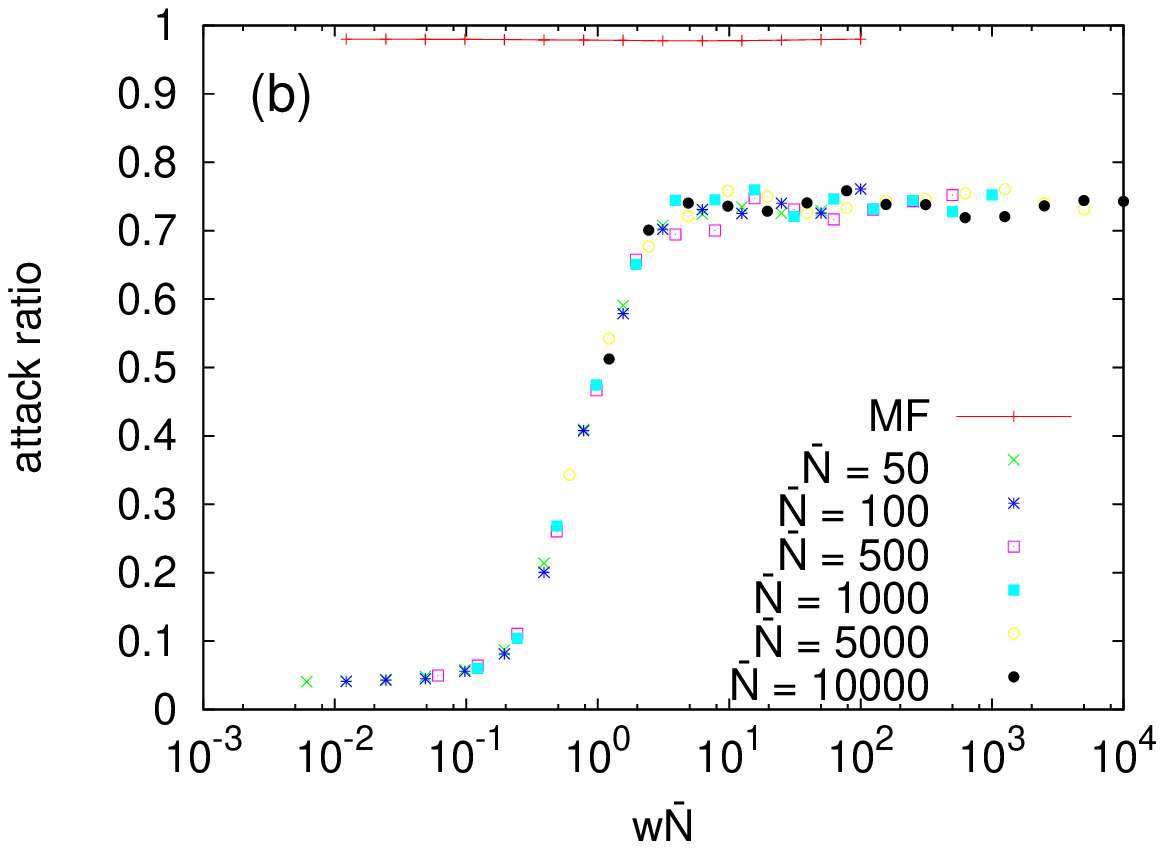}
\caption{Final size distributions as functions of $w\bar{N}$ for (a) a
 metapopulation SIR model and (b) a metapopulation SHIR model. 
 The transmission rate is $\beta=1$.
 Monte Carlo simulations for $\bar{N}=50$, $100$, $500$, $1000$, $5000$,
 and $10000$  are
 compared with the result in a reaction-diffusion system, namely, a
 mean-field (MF) approach.}
\label{f2s}
\end{figure}

Moreover, in Monte Carlo simulations, the attack ratio depends not
only on the mobility rate $w$ but 
also on the average population $\bar{N}$ of a patch.
The quantity that defines the global invasion threshold is proportional
to $w\bar{N}$, and depends on the network structure and other
parameters~\cite{Colizza07}.  
In Fig.~\ref{f2s}, the attack ratio is plotted as a function of $w\bar{N}$
for $\beta=1$. For different values of $\bar{N}$, the attack ratio
simulated with the Monte Carlo method collapses into a single curve. 
The attack ratio increases rapidly at around  
$w\bar{N}\sim 1$ and saturates to a certain value.
The saturated value of attack ratio in the Monte Carlo simulations is
approximately $0.7$--$0.8$, while that in the mean-field approach is almost $1$.
This is because infected individuals disappear at some probability
before the epidemic grows. 
This type of behavior is often observed when the initial condition has
only one infected individual in the entire metapopulation. 

\begin{figure}[tbh]
\includegraphics[width=8cm]{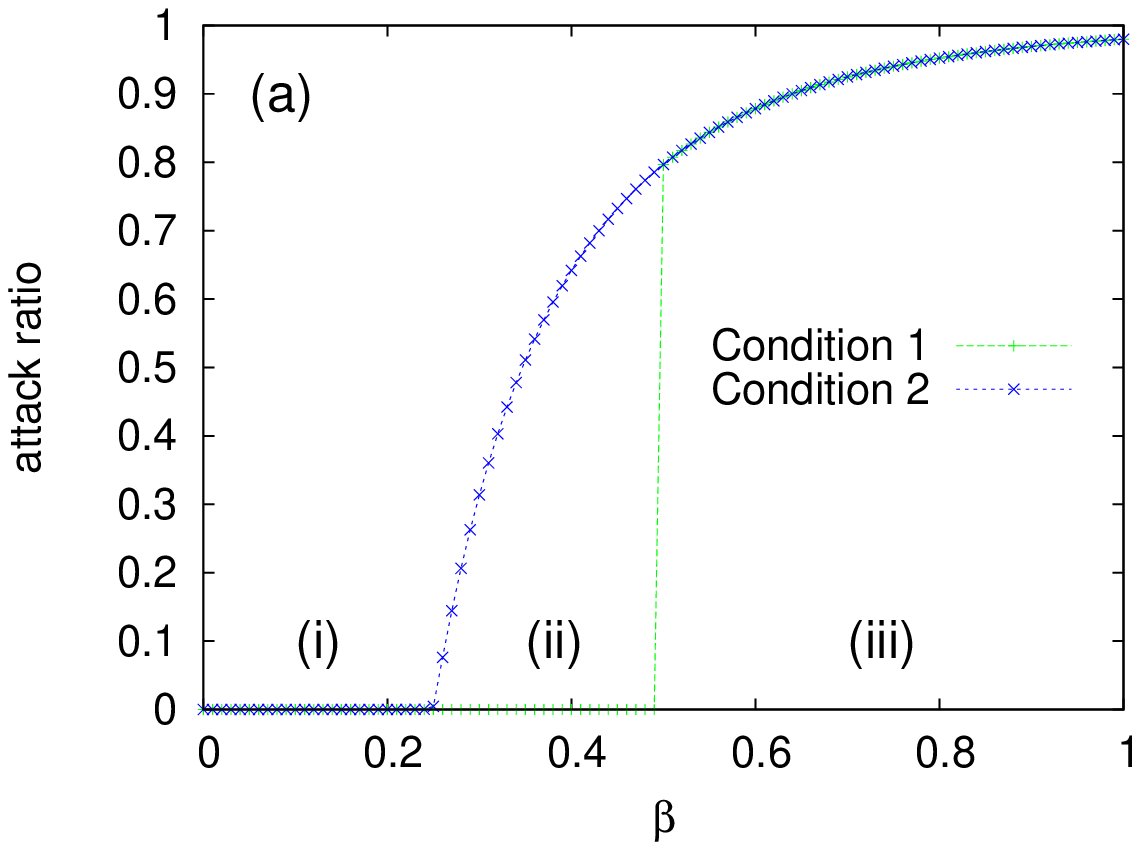}
\includegraphics[width=8cm]{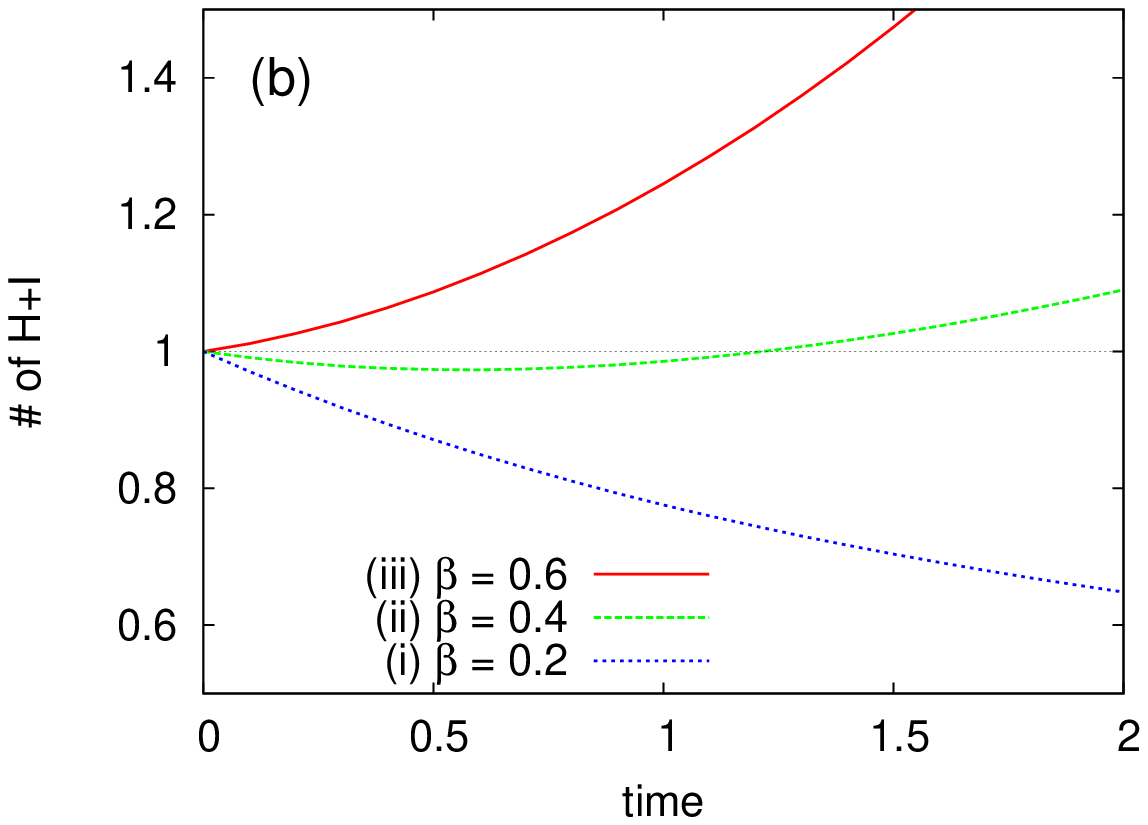}
\caption{(a) Final size distributions for the metapopulation SHIR model with
 $w=1$. Termination conditions are $\sum_i (H_i+I_i) < 1$ for
 Condition 1 and $\sum_i (H_i+I_i)=0$
 for Condition 2. Interval $0<\beta<1$ is separated into three regimes: 
(i) $0<\beta<0.25$, (ii) $0.25<\beta<0.5$, and (iii) $0.5<\beta<1$. 
(b) Time evolution of the number of infected and seriously-ill
 individuals $\sum_i(H_i+I_i)$. $\beta=0.2, 0.4$ and $0.6$ are
 representative points in Regimes (i), (ii), and (iii), respectively.} 
\label{f3}
\end{figure}

Here, we explain the reason for the discontinuous jump in the 
reaction-diffusion system.
The key is the termination condition.
When the termination threshold is almost zero, the
discontinuous jump disappears. 
The final size distributions under different termination conditions are
shown in Fig.~\ref{f3}(a), where the mobility rate is fixed at $w=1$.
The green curve in Fig.~\ref{f3}(a) is the same as the final size
distribution in
Fig.~\ref{f2}(b). The blue curve in Fig.~\ref{f3}(a) is calculated under
the termination condition $\sum_i (H_i+I_i) =0$ (more precisely,
$\sum_i (H_i+I_i) < \epsilon$ with $\epsilon= 10^{-14}$).
It has a continuous transition at $\beta=0.25$, which corresponds to
$R_0=1$. 
To understand the effects of different termination conditions, we
separate the interval of $\beta$ into three regimes:
(i) $0<\beta<0.25$, (ii) $0.25<\beta<0.5$, and (iii) $0.5<\beta<1$. 
In Regime (i), the number of infected individuals decreases
monotonically as shown in Fig.~\ref{f3}(b).
Since $R_0$ is lower than the epidemic threshold, a disease cannot
spread at all.
By contrast, in Regime (iii), initially, the number of infected individuals
increases rapidly.
In these two regimes, termination conditions have no influence on the attack
ratio. However, in Regime (ii), the situation is a bit complicated.
The number of infected individuals decreases at first but increases
after a short interval of time. This initial decay in $\sum_i (H_i+I_i)$ is natural,
since 
\begin{equation}
 \frac{d}{dt} \sum_i (H_i+I_i) = \sum_i\left[
\beta \frac{S_i(H_i+I_i)}{N_i} -\gamma I_i \right]
\simeq (\beta - \gamma)\sum_i (H_i+I_i)
\end{equation}
with $H_i=0$ and $S_i\simeq N_i$ at $t=0$.
Since $\beta<\gamma$ in Regime (ii), the simulation
stops immediately when the termination threshold is $1$. 
When the threshold is closer to 0 (or $\epsilon$), 
the simulation continues, and the number of infected
individuals grows enough since $R_0>1$. 
In the SIR model, $\beta=\gamma$ is equivalent to $R_0=1$. Thus, Regime
(ii) does not exist in the SIR model.

\section{\label{sec:4} Conclusions}

We introduced a modified metapopulation SIR model with a latent
period, which we 
call the SHIR model, and compared the final size distributions of the
metapopulation SIR and SHIR models. 
In Monte Carlo simulations, the distributions of both the models appear similar and
depend on the mobility rate as well as the transmission rate.
However, those in the reaction-diffusion systems are independent of the 
mobility rate and have different properties between the SIR and SHIR model. 
The difference between the two methods arises from the treatment of
population; population is treated as an integer in Monte Carlo
simulations and a real number in reaction-diffusion systems.
Although the final size distribution of the SIR model has a continuous
transition, that of the SHIR model shows a discontinuous jump,
depending on the termination condition of the simulation.
The difference caused by the termination condition indicates that a
temporal decay of the number of infected individuals can mislead us into
underestimating the final size of an epidemic.

\end{document}